\documentclass[aps,prl,twocolumn,showpacs,superscriptaddress,groupedaddress]{revtex4}  

\usepackage{graphicx}  
\usepackage{dcolumn}  
\usepackage{bm}
\usepackage{amsmath}
\usepackage{amssymb}
\usepackage{amsthm}
\usepackage{amsfonts}
\usepackage{listings}

\newcommand{\beq}{\begin{equation}}
\newcommand{\eneq}{\end{equation}}

\newcommand{\bra}[1]{\left\langle#1\right|}
\newcommand{\ket}[1]{\left|#1\right\rangle}

\def\be{\begin{equation}}
\def\ee{\end{equation}}
\def\ba{\begin{eqnarray}}
\def\ea{\end{eqnarray}}

\begin{document}

\title{$D$-Algebra Structure of Topological Insulators}

\author{B. Estienne}
\affiliation{Department of Physics, Princeton University, Princeton, NJ 08544}
\author{N. Regnault}
\affiliation{Department of Physics, Princeton University, Princeton, NJ 08544}
\affiliation{Laboratoire Pierre Aigrain, ENS and CNRS, 24 Rue Lhomond, 75005 Paris, France}
\author{B. A. Bernevig}
\affiliation{Department of Physics, Princeton University, Princeton, NJ 08544}

\date{\today}

\begin{abstract}
In the quantum Hall effect, the density operators at different wave-vectors generally do not commute and give rise to the Girvin MacDonald Plazmann (GMP) algebra with important consequences such as ground-state center of mass degeneracy at fractional filling fraction, and $W_{1 + \infty}$ symmetry of the filled Landau levels.  We show that the natural generalization of the GMP  algebra to higher dimensional topological insulators involves the concept of a $D$-commutator. For insulators in even dimensional space, the $D$-commutator is isotropic and closes, and its structure factors are proportional to the $D/2$-Chern number. In odd dimensions, the algebra is not isotropic, contains the weak topological insulator index (layers of the topological insulator in one fewer dimension) and does \emph{not} contain the Chern-Simons $\theta$ form.   This algebraic structure paves the way towards the identification of fractional topological insulators through the counting of their excitations.  The possible relation to $D$-dimensional volume preserving diffeomorphisms and parallel transport of extended objects  is also discussed.
\end{abstract}

\maketitle

{\bf Introduction:} Fractional topological insulators (FTI) are the strongly correlated states that may appear when a narrow bandwidth bulk band of a topological insulator~\cite{kane-PhysRevLett.95.226801,Bernevig15122006} is fractionally filled and subject to strong interactions.   Evidence for their existence has been provided in a series of analytical and numerical works in two-dimensional Chern insulators\cite{sheng-natcommun.2.389,regnault-PhysRevX.1.021014,neupert-PhysRevLett.106.236804,wang-PhysRevLett.107.146803,wu-PhysRevB.85.075116,qi-PhysRevLett.107.126803,wang-PhysRevB.84.241103,McGreevy-PhysRevB.85.125105,jiang-PhysRevB.84.205324} and time-reversal invariant topological insulators \cite{neupert-PhysRevB.84.165107,neupert-PhysRevB.84.165138,Taylor-inprep2012}.  The plethora of new experimental facts and theoretical ideas discovered in the non-interacting topological insulators suggests that their fractional (\emph{i.e.} interacting) counterparts will also exhibit new physical properties of topological phases, especially in space dimensions higher than two.

The excitation counting of a topological state of matter is an imprint of the the underlying topological phase. It contains information about the  nature of both the excitations and the edge states.
The most studied fractional topological insulator, the two-dimensional fractional Chern insulator (FCI), has been identified through the counting of its excitations  (in both the energy and the entanglement spectrum) \cite{Bernevig-2012PhysRevB.85.075128}. 
This progress was made possible by the non trivial algebra obeyed by its projected density operators \cite{Parameswaran-PhysRevB.85.241308}. For a smooth enough  Berry curvature  in the Brillouin zone (BZ), this algebra is nothing but the celebrated  Girvin-MacDonald-Plazmann (GMP)  algebra of the  fractional quantum Hall (FQH) effect \cite{girvin-PhysRevB.33.2481}.
 This algebra has far reaching consequences: it is identical to the algebra of area-preserving diffeomorphisms, thereby providing for an explanation of the edge modes of an integer quantum Hall liquid as shape deformations of the liquid droplet. It allows for the construction of nontrivial many-body symmetry operators of the Hilbert space, it provides for a center of mass degeneracy (exact in the FQH but approximate in the FCI), and is related to the Hall viscosity, $q^4$ form factor, as well as the edge dipole-moment \cite{Wiegmann-PhysRevLett.108.206810}.

All numerical studies of FTI  in higher dimensions rely on excitation counting as conclusive evidence. A prerequisite is to develop analytical tools that apply to dimensions greater than two.    In this Rapid Communication, we present  a generalization of the GMP algebra to topological insulators in higher dimensions.  In even space dimensions we consider Chern insulators (A class\cite{Kitaev-2009AIPC.1134.22K,Ryu-2010NJPh.12f5010R}), which are characterized by a Chern number. In odd dimensions we consider $\mathbb{Z}_2$ topological insulators, whose topological number is the average over the Brillouin zone of the Chern-Simons form. We generalize the usual commutator structure in $D$ spatial dimensions to a $D$-commutator by contracting with the antisymmetric tensor in $D$-dimensions.  If this commutator is closed, the relation is called a $D$-algebra. We find that for topological insulators in even dimensions, the commutator is closed, and the algebra is isotropic, under a condition similar to that of the existence of the GMP in the $2$-D Chern insulator\cite{Parameswaran-PhysRevB.85.241308}. Its structure factors are proportional to the $D/2$'th Chern number. In odd space dimensions, however, the density algebra does not probe the relevant topological number, as the Chern-Simons form ($F\wedge A +i/3 A\wedge A \wedge A$ in 3D) does not appear. This algebra is anisotropic in nature, as it is sensitive to layers of $(D-1)$-dimensional topological insulators in the system.  This algebraic structure opens a way towards the identification of fractional topological insulators through the counting of their excitations.

{\bf  Projected density operators and flat-band limit:} We start by fixing notations and recalling some well known results about band structure and projected density operators for topological insulators. We consider a $N$ band topological insulator described by a translationally invariant Hamiltonian, and we work on a $D$ dimensional lattice  (with $L^D$ sites) with periodic boundary conditions. After diagonalization of the the Bloch matrix, the one-body Hamiltonian takes the form
\begin{equation}
 H = \sum_{{\bf k}, n} E_n({\bf k}) \gamma_{{\bf k}}^{n \dagger} \gamma_{{\bf k}}^{n}, \label{translationalinvariance}
\end{equation}
where the normal modes $\gamma_{\bf k}^n$ can be written as a matrix rotation of the original electron operators $\gamma_{{\bf k}}^n = \sum_\beta u^{n \star}_{{\bf k}, \beta} c_{{\bf k}, \beta}$.
We consider the physics of the (possibly fractionally) occupied bands and look only at projectors into these bands. The projection operator in the occupied bands is  
$ P =   \sum_{n,{\bf k} }  | {\bf k},n \rangle \langle n,{\bf k} | $
where $|n, {\bf k}\rangle = \gamma^{n \dagger}_{\bf k}|0 \rangle$ and the band index $n$ ranges over all occupied bands $n= 1, \cdots, N_{\text{occ}}$ . 
The density operator $e^{- i {\bf q}\cdot {\bf r}} = \sum_{{\bf j}, \alpha} e^{-i {\bf q}\cdot {\bf j}} c_{{\bf j}\alpha}^\dagger c_{{\bf j} \alpha}$ becomes when projected to the occupied bands of a topological insulator:
\begin{equation}
 \rho_{{\bf q}}  = \sum_{{\bf k}, n,m} \langle u_{\bf k}^n | u_{{\bf k}+{\bf q}}^{m} \rangle \gamma^{n \dagger}_{{\bf k}} \ket{0} \bra{0} \gamma^{m}_{{\bf k} +{ \bf q}},
 \label{lowmomentumexpansion}
\end{equation}
where $n,m$ range over the set of occupied bands. 

Fractional topological insulators are usually constructed and observed in models with fractionally filled bands whose bandwidth is very small, such that interactions and not the kinetic energy dominate the physics. The ideal example of such an insulator is the flat-band model, which gives an energy $-1$ to occupied bands, and $+1$ otherwise
\begin{align}
H_{FB}=1-2P \, .\label{FB}
\end{align}
All projected operators commute with the this deformed one-body Hamiltonian. Therefore projected density operators are an exact symmetry of the flat-band Hamiltonian, to which the true one-body Hamiltonian (\ref{translationalinvariance}) is adiabatically connected.

{\bf Two-dimensional algebra and first Chern number:} Before moving to higher dimensions, we quickly review what is known about the algebra of projected density operators in two dimensions, with an emphasis to its main characteristics. We focus on the appearance of the Chern number in the algebra and on the link between projected densities and parallel transport in the background of the Berry curvature.  At long wavelength (${\bf q}_{1},{\bf q}_{2} \rightarrow 0$), in Ref.~\cite{Parameswaran-PhysRevB.85.241308} the following commutation relation was found:

\beq
[\rho_{{\bf q}_1}, \rho_{{\bf q}_2}]=- i q_{1}^{\mu} q_{2}^{\nu} \sum_{{\bf k}, n,m} F_{\mu \nu}^{n, m}({\bf k})  \gamma_{\bf k}^{n \dagger} \ket{0}{\bra{0}} \gamma_{{\bf k}+{\bf q}_1+{\bf q}_2}^{m} , \label{2-commutator}
\eneq
 where  the Einstein summation convention over repeated indices is assumed.  This result holds in any dimension. $F_{\mu \nu} = \partial_{\mu} A_{\nu} - \partial_{\nu} A_{\mu} - i [A_{\mu},A_{\nu}] $ is the non-Abelian Berry field strength in the Brillouin zone, while the vector potential is $A^{nm}_{\mu}({\bf k}) =  i \langle u_{\bf k}^n | \partial_{k_{\mu}} | u_{\bf k}^m \rangle $.

In two dimensions $F_{\mu \nu}({\bf k})= B ({\bf k})\epsilon_{\mu \nu}$, and its integral over the whole Brillouin zone yields the first Chern number $
C_1 =  \frac{1}{4\pi}\int_{\text{BZ}} d^2k \epsilon^{\mu \nu} \text{Tr} \left( F_{\mu \nu}({\bf k}) \right)$ .
The commutator of two densities has to be non-zero in a nontrivial Chern insulator. That is so because the Chern number $C_1$ of the two-dimensional  insulator can be expressed as a trace over the Brillouin zone of the density commutator 
\begin{align}
\text{Tr}\left([\rho_{{\bf q}_1}, \rho_{{\bf q}_2}]\rho_{-{\bf q}_1 - {\bf q}_2}) \right) \sim_{{\bf q} \to 0} \frac{L^2}{ 2\pi i} \, ({\bf q}_1 \wedge {\bf q}_2) \, C_1 \label{ChernTrace2D}
\end{align}
where ${\bf q}_1 \wedge {\bf q}_2 = \epsilon_{\mu \nu} q_{1}^{\mu} q_{2}^{\nu}$.
In the continuum limit of the quantum Hall effect, the projected density algebra of the Lowest Landau Level  is called the GMP algebra. Its generators are the generators of the area-preserving diffeomorphisms in two-dimensions. This result is recovered for two-dimensional topological insulators with an Abelian $\text{U}(1)$ uniform Berry curvature,  in the long wavelength limit.  As pointed out in Ref.~\cite{Parameswaran-PhysRevB.85.241308} (see also Refs.~\cite{goerbig-2012epjb} and \cite{Bernevig-2012PhysRevB.85.075128}),  if the local Berry curvature can be replaced by its average $F_{xy} ({\bf k}) = B = \frac{C_1}{2\pi} $ then $[\rho_{{\bf q}_1}, \rho_{{\bf q}_2}]    = -  iB \, {\bf q}_1 \wedge {\bf q}_2 \,  \rho_{{\bf q}_1 + {\bf q}_2} $.
Note that ${\bf q}_1 \wedge {\bf q}_2$  is the area enclosed in the parallelogram delimited by ${\bf q}_1$ and ${\bf q}_2$. This algebra is nothing but the two-dimensional Aharonov-Bohm effect in momentum space, in the background of the "magnetic field" $F_{x y} = B$. Expanding the projected densities at long wave-vectors  as $\rho_{\bf q} = 1 + i {\bf q}\cdot {\bf R} + O({\bf q}^2)$, the algebra of the guiding center is recovered 
\begin{align}
[R_1,R_2] = i B = \frac{i}{2\pi}C_1  \, . \label{14}
\end{align}
The Chern number quantifies the non-commutativity of the guiding center operators.  This Abelian treatment applies to two-band models (insulators with one band below and above the gap) or to many-band insulators where the non-Abelian components of the field strength can be neglected (up to an overall prefactor $N_{occ}$). We remark that in a two-band insulator, it is impossible to have a constant Berry curvature due to the no-hair theorem\cite{podolskiandavronprivatecommunication}, although this seems possible in insulators with four or more bands \cite{podolskiandavronprivatecommunication}.

Since projected density operators commute with the flat-band Hamiltonian \eqref{FB},  it would seem that they are the generators of a proper symmetry group of the system. However this is not quite true, as they suffer from a serious deficiency. Because of the projection, they are not unitary.  The density operator translates states in momentum space but does not keep their norm:
\beq
\rho_{{\bf q}} \ket{n, {\bf k}} =  \sum_{m} \langle u^{m }_{{\bf k}-{\bf q}} | u^{n}_{{\bf k}} \rangle \ket{m, {\bf k}-{\bf q}} \, . \label{momentum translation}
\eneq
It is possible to replace the projected density operator $\rho_{{\bf q}}$ by a unitary operator $\tilde{\rho}_{{\bf q}}$ , while not spoiling the long wavelength behavior from Eq.\eqref{2-commutator}. For a uniform Abelian Berry curvature, the answer is quite straightforward, and is simply the exponentiation of the guiding center operator. Doing this, one recovers the GMP algebra
\begin{align}
[\tilde{\rho}_{{\bf q}_1}, \tilde{\rho}_{{\bf q}_2}]  &  = -2 i\sin \left(  B \frac{ {\bf q}_1 \wedge {\bf q}_2}{2}\right)  \tilde{\rho}_{{\bf q}_1 + {\bf q}_2} \, .\label{regularizeddensityalgebra4}
\end{align}
 More generally, for a non-Abelian and non-uniform Berry field strength, the answer is parallel transport in the background of the Berry gauge potential $A_{\mu}({\bf k})$ :
\beq
\tilde{\rho}_{{\bf q}}=\sum_{{\bf k}; n,m} \left( \mathcal{P} e^{-i \int_{{\bf k}}^{{\bf k}+{\bf q}} {A}({\bf k}') dk'}   \right)_{nm} \gamma_{{\bf k}}^{n \dagger} \ket{0}{\bra{0}} \gamma_{{\bf k}+{\bf q}}^{m} \label{regularized density operator} \, .
\eneq 
 In the Abelian case this result was pointed out in Ref.~\cite{Parameswaran-PhysRevB.85.241308}. Note that the parallel transport also commutes with the Flat-Band Hamiltonian, and at small momenta coincides with the projected density operator  $ \tilde{\rho}_{\bf q} = \rho_{\bf q}  + O({\bf q}^2)$. 

{\bf Density algebra in even-space dimensions: } The density commutator is natural in two dimensions. In higher space dimension $D>2$, the commutator algebra Eq.\eqref{2-commutator} reveals whether a two dimensional quantum Hall effect exists on a given plane of the $D$-dimensional space defined by the two vectors ${\bf q}_1$ and ${\bf q}_2$. It is apparent then that the commutator algebra Eq.\eqref{2-commutator} cannot probe isotropic quantities such as the topological number. In order to find an isotropic algebraic structure in higher dimensions, we must look somewhere else. We first realize that the commutator $[\rho_{{\bf q}_1}, \rho_{{\bf q}_2}]$ is, in two dimensions, simply a re-writing of the operators $\epsilon_{\alpha \beta} \rho_{{\bf q}_\alpha} \rho_{{\bf q}_\beta}$. In $D$ space dimensions, it is then suggestive to look at the operator
\begin{align}
 \left[\rho_{{\bf q}_{\alpha_1}}, \rho_{{\bf q}_{\alpha_2}},\cdots, \rho_{{\bf q}_{\alpha_D}}\right] = \epsilon_{\alpha_1 \alpha_2 \cdots \alpha_D} \rho_{{\bf q}_{\alpha_1}} \rho_{{\bf q}_{\alpha_2}} \cdots \rho_{{\bf q}_{\alpha_D}},
\end{align}
where $\epsilon_{\alpha_1, \alpha_2, \ldots, \alpha_D} $ is the totally antisymmetric tensor in $D$-dimensions. and $\alpha =1\ldots D$. These generalized commutators are called $D$-commutators.  We will now compute this object in the longe wavelength limit and find it is closed, thereby generating a $D$-algebra.

The density algebra in even space dimensions is simpler to obtain than in odd-space dimensions for reasons that will become apparent. In even space dimensions we have the Chern-insulator (QH)-classes, so we anticipate that the algebra closes. We first re-express the $D$-commutator as a product of $2$-commutators:  $ \left[\rho_{{\bf q}_{1}},\cdots, \rho_{{\bf q}_{D}}\right] = 2^{-D/2} \epsilon_{\alpha_1 \ldots  \alpha_D}  [\rho_{{\bf q}_{\alpha_1}},  \rho_{{\bf q}_{\alpha_2}}] \cdots [ \rho_{{\bf q}_{\alpha_{D-1}}}, \rho_{{\bf q}_{\alpha_D}} ]$. Using the long wavelength two dimensional algebra Eq.\eqref{2-commutator} and working at order ${\bf q}^D$ we obtained :
\begin{align}
& \left[\rho_{{\bf q}_{1}}, \rho_{{\bf q}_{2}},\cdots, \rho_{{\bf q}_{D}}\right]  = (-i)^{D/2}  \left({\bf q}_1 \wedge {\bf q}_2 \wedge \cdots \wedge {\bf q}_D \right) \times \nonumber \\ & \sum_{{\bf k}, n, m} \left( F({\bf k}) \wedge \cdots  \wedge F({\bf k})  \right)_{n m} \gamma_{{\bf k}}^{n \dagger} \ket{0}\bra{0} \gamma^{m}_{{\bf k} +{\bf q}_{1} + \cdots +   {\bf q}_{D}} . \label{EvenDalgebra}
\end{align}
This equation is the $D$-dimensional analogue of Eq.\eqref{2-commutator}. In the $D$-commutator appear the matrix $ F \wedge \cdots  \wedge F  = 2^{-D/2} \epsilon^{\mu_1  \cdots  \mu_D} F_{\mu_1 \mu_2} \cdots  F_{\mu_{D-1} \mu_D}$
 which is the $D/2$'th Berry curvature density of the $D/2$'th Chern number:
\beq
C_{D/2} = \frac{1}{(D/2)! (2\pi)^{D/2}} \int d^D k \text{Tr}\left( F({\bf k}) \wedge \cdots  \wedge F({\bf k}) \right). \label{Chern_trace_Deven}
\eneq
 For even dimensional topological insulators, the  $D/2$'th Chern number can be expressed as the the trace over the $D$-commutator of the projected density operator:
\begin{align}
 \text{Tr} \left( [\rho_{{\bf q}_{1}}, \rho_{{\bf q}_{2}},\cdots, \rho_{{\bf q}_{D}}] \rho_{-({\bf q}_1 + \ldots + {\bf q}_D) } \right) \nonumber \sim_{{\bf q} \to 0}\\  \frac{L^D}{(2\pi i)^{D/2}} (D/2)! \left({\bf q}_1 \wedge {\bf q}_2 \wedge \cdots \wedge {\bf q}_D \right)  C_{D/2} \label{ChernTraceEvenD}.
\end{align}
This is the exact analog of the two-dimensional relation \eqref{ChernTrace2D}.

It is possible to obtain an analog of the GMP algebra in $D$-dimensions.  As for topological insulators in two dimensions,  this algebra holds when the Berry density $ F({\bf k}) \wedge \cdots  \wedge F({\bf k})$  is uniform in the Brillouin zone, and proportional to the identity matrix. This situation is not as restrictive as it may seem, and we conjecture that Chern insulators are adiabatically connected to this uniform case.  For instance the integer quantum Hall effect in $2,4$ and $8$ dimensions \cite{Zhang-2001Sci.294.823Z,PhysRevLett.91.236803} enjoy these properties, as inherited from the underlying monopole field configurations. Under these assumptions the  projected density operators algebra closes in the long wavelength limit 
\begin{align}
\left[\rho_{{\bf q}_{1}}, \rho_{{\bf q}_{2}},\cdots, \rho_{{\bf q}_{D}}\right]  =   (D/2)!\frac{1}{(2\pi i)^{D/2}} \frac{C_{D/2}}{N_{occ}} \nonumber \\   \left( {\bf q}_1 \wedge {\bf q}_2 \wedge \cdots \wedge {\bf q}_D \right)  \rho_{{\bf q}_1 +\ldots + {\bf q}_D },
 \label{D-commutator even}
\end{align}  and we recover a $D$-algebra. It is very tempting to expand the projected densities as $\rho_{\bf q} = 1 + i {\bf q}\cdot {\bf R} + O({\bf q}^2)$. The "guiding center"  algebra is most easily obtained in the continuum limit. From Eq.\eqref{momentum translation} the guiding center operators are simply the covariant derivative  with the Berry potential in momentum space $R_{\mu}({\bf k}) =- i \left( \partial_{k_\mu} - iA_{\mu}({\bf k}) \right) $.
Using the relation $[R_{\mu},R_{\nu}] = i F_{\mu \nu}$, it is straightforward to obtain their $D$-commutator
\begin{align}
[R_{1}({\bf k}),\cdots, R_{D}({\bf k})] = i^{D/2} F({\bf k}) \wedge \cdots  \wedge F({\bf k}) . \label{D guiding center}
\end{align}
This elementary derivation in the continuum is not plagued by the limitations of derivation on the lattice, as there is no need to suppose the Berry density  $ F({\bf k}) \wedge \cdots  \wedge F({\bf k})$   to be uniform or proportional to the identity.

This $D$-algebra structure may be understood in two ways. On one hand, as was pointed out in Ref.~\cite{Bernevig-2012PhysRevB.85.075128}, the projected position operators can be expressed in terms of the projected density operators.  Therefore an immediate interpretation of  Eq.\eqref{D guiding center} is the non commutativity of the coordinates of particles projected to the occupied bands of a topological insulator.  This is the $D$-dimensional analog of Eq.\eqref{14}  for the Quantum Hall Effect.

On the other hand, the GMP algebra also describes a two-dimensional Aharonov-Bohm effect: the projected density operators implement parallel transport of pointlike objects in the background of the Berry curvature $F$. In higher dimensions,  an Aharonov-Bohm effect with respect to the $D$-form $F\wedge \cdots \wedge F$  requires parallel transport of higher dimensional objects. 
 We conjecture that the algebra \eqref{D guiding center} is related to an Aharonov-Bohm effect involving extended excitations (membranes) coupled to the Berry curvature $F\wedge F\wedge \cdots \wedge F$. However, unlike in two dimensions, it is not clear how to interpret the projected density operators as an implementation of membrane parallel transport.

{\bf Density algebra in odd-space dimensions: } Pursuing the same strategy in odd dimensions leads to an impasse. The topological invariant in odd dimensions is defined as the integral over the Brillouin zone of a Chern-Simons form. For instance in three dimensions  the $\mathbb{Z}_2$ topological invariant is given by
\begin{align}
P_3 = \frac{\theta}{2\pi} = \frac{1}{8 \pi^2} \int d^3k \, \text{Tr} \left[ F \wedge A + \frac{i}{3} A \wedge A \wedge  A \right]  . \label{CSterm}
\end{align}
Defined for all odd dimensions, a characteristic feature of Chern-Simons form is that their integral is not invariant under large gauge transformations. However the variation has to be an integer  \cite{Dunne1990281}. In contrast to the even dimensional Chern numbers, the odd dimensional $\mathbb{Z}_2$  topological invariant is gauge invariant only modulo integers.  Trying to obtain $P_3$  through the gauge invariant trace $\text{Tr}([\rho_{{\bf q}_1}, \rho_{{\bf q}_2}, \rho_{{\bf q}_3}]\rho_{-{\bf q}_1 - {\bf q}_2 - {\bf q}_3})$ is doomed to fail. A simple relation like Eq.\eqref{ChernTraceEvenD} is ruled out in odd dimensions. Moreover  $D$-commutators in odd dimensions are known \cite{Curtright-PhysRevD.68.085001} be  more problematic than their even dimensional counterparts. For instance while even commutators involving the identity matrix do vanish, this is no longer the case for 
odd commutators. This is most easily seen in $3$ dimensions:
\begin{align}
[A,B,1] = [A,B] \neq 0  .
\end{align}
Consequently, when expanding the $3$-commutator of a projected density operator $\rho_{\bf q} = 1 + i {\bf q}\cdot {\bf R} + O({\bf q}^2)$, the lowest order contribution is of order $q^2$ and not $q^3$:
\begin{align}
[\rho_{{\bf q}_1}, \rho_{{\bf q}_2}, \rho_{{\bf q}_3}] \sim  - i (q_1^{\mu}q_2^{\nu} +q_3^{\mu}q_1^{\nu} + q_2^{\mu}q_3^{\nu} ) F_{\mu \nu} \rho_{{\bf q}_1+{\bf q}_2+{\bf q}_3} . \label{3commutatorq^2}
\end{align}
This term is reminiscent of the 2-commutator algebra Eq.\eqref{2-commutator}, and accounts for a possible two-dimensional topological structure in the 3D insulator. This would be the case for a weak 3D Chern insulator, obtained by stacking layers of the 2D Chern insulator. This structure remains true in \emph{all} odd dimensions, where the $D$-commutator contains an anisotropic $O({\bf q}^{D-1})$ term in contrast with the isotropic $O({\bf q}^D)$  term appearing in Eq.\eqref{EvenDalgebra} for even dimensions.

In order to investigate in more detail  the kind of problems that arise in odd dimensions, we computed the sub-leading term in the algebra \eqref{3commutatorq^2} in three dimensions. If the Chern-Simons density \eqref{CSterm} is to appear at all in the triple commutator, this has to be as a $O({\bf q}^3)$ term. Upon computing the sub-leading term of the $3$-commutator $[\rho_{{\bf q}_1}, \rho_{{\bf q}_2}, \rho_{{\bf q}_3}]$, a term $ ({\bf q}_1 \wedge {\bf q}_2 \wedge {\bf q}_3 ) F \wedge A$ appears.  This promising term is part of the Chern-Simons form, although the $\frac{i}{3}A \wedge A \wedge A$ part is missing. However in order to close the algebra we need to multiply the $3$-commutator by $\rho_{-{\bf q}_1 - {\bf q}_2 - {\bf q}_3}$, and rather than completing the Chern-Simons term, it kills it altogether. We are left with
\begin{align}
[\rho_{{\bf q}_1}, \rho_{{\bf q}_2}, \rho_{{\bf q}_3}]\rho_{-{\bf q}_1 - {\bf q}_2 - {\bf q}_3}  = \nonumber \\
 -i \sum_{{\bf k},n,m} (q_1^{\mu}q_2^{\nu}+ q_2^{\mu}q_3^{\nu}+q_3^{\mu}q_1^{\nu})(F_{\mu \nu})_{nm} \gamma_{{\bf k}}^{n \dagger} \ket{0}\bra{0} \gamma^{m}_{{\bf k} } \nonumber \\  
+  \epsilon_{\alpha_1 \alpha_2 \alpha_3}q^{\mu}_{\alpha_1}q^{\nu}_{\alpha_1}  q_{\alpha_2}^{\sigma} \frac{1}{2}\sum_{{\bf k},n,m}  \left(C_{\mu \nu \sigma}\right)_{nm}  \gamma_{{\bf k}}^{n \dagger} \ket{0}\bra{0} \gamma^{m}_{{\bf k}} 
\end{align}
The sub-leading term does not contain the expected antisymmetric tensor $ ({\bf q}_1 \wedge {\bf q}_2 \wedge {\bf q}_3 ) \epsilon^{\mu \nu \sigma}$. Instead we have the tensor $\epsilon_{\alpha_1 \alpha_2 \alpha_3}q^{\mu}_{\alpha_1}q^{\nu}_{\alpha_1}  q_{\alpha_2}^{\sigma}$, which is symmetric under $\mu \leftrightarrow \nu$, and cannot be contracted to the antisymmetric Chern-Simons tensor. Instead it comes with the tensor 
\begin{align}
C_{\mu \nu \sigma} & = iD_{\sigma}B_{\mu \nu} -i \partial_{\mu} \partial_{\nu}A_{\sigma}   -  (A_{\mu}\partial_{\nu} + A_{\nu}\partial_{\mu})A_{\sigma} \nonumber \\ & +  F_{\mu \sigma}A_{\nu} +F_{\nu \sigma}A_{\mu}, 
\end{align} 
where $D_{\sigma}\cdot = \partial_{\sigma} \cdot+ i [A_{\sigma},\cdot ]$ and $B_{\mu \nu}$ is the $O({\bf q}^2)$ regularization of the density operator $\rho_{\bf q}  = \sum_{{\bf k},n,m} \left( 1 - i q^{\mu}  A_{\mu} - \frac{i}{2} q^{\mu} q^{\nu}B_{\mu \nu} \right)_{nm} \gamma_{{\bf k}}^{n \dagger} \ket{0}{\bra{0}} \gamma_{{\bf k}+{\bf q}}^{m} $. 
The $3$-tensor $C_{\mu \nu \sigma}$ being $\mu \leftrightarrow \nu$ symmetric, it can never yield the fully antisymmetric Chern-Simons term, and this calculation shows explicitly that the Berry curvature does appear in the algebra of projected density operators in three dimensions, no matter what regularization $B_{\mu \nu}$ is chosen for the density operator.

A way to get around this no-go theorem is to involve non gauge invariant operators, such as the pure translation $T_{\bf q}  \ket{n , {\bf k}} =    \ket{n, {\bf k}-{\bf q}}$. This can be used to generate the Chern-Simons form as
\begin{align}
P_3 = \frac{1}{32 \pi ^2}\epsilon_{ijk} \text{Tr}\left[    (  \rho_{{\bf q}_i} \rho_{{\bf q}_j} (\rho_{{\bf q}_k} - T_{{\bf q}_k} )- \right. \nonumber \\ \left. \frac{1}{3} ( \rho_{{\bf q}_i} - I_{{\bf q}_i} )( \rho_{{\bf q}_j}  - T_{{\bf q}_j} )( \rho_{{\bf q}_k} - T_{{\bf q}_k} )) \rho_{- {\bf q}_i - {\bf q}_j - {\bf q}_k} \right],
\end{align}
but the  physical picture behind this relation is still unclear.\\
{ \bf Concluding remarks:} We have presented a generalization of the  GMP algebra to $D$-dimensional topological insulators by generalizing the commutator, algebra and Berry phase to their higher-dimensional counterparts.  At this level, the even and odd-dimensions are fundamentally different - in even dimensions, the structure factors of the algebra are proportional to the $D/2$'th Chern number, while in odd dimensions they are not proportional to the expected Chern-Simons form. The $D$-commutator hints at a different group structure from the usual gauge theories, such as higher gauge theories \cite{baez-commmathphys.293.2010,baez-2011GReGr..43.2335B}. In light of this, the recent proposal \cite{cho-2011AnPhy.326.1515C} to describe topological insulators by a BF theory \cite{Horowitz-CommMathPhys.1989} looks very promising. In two dimensions, the classical limit of the GMP algebra is isomorphic to the algebra of area preserving diffeomorphisms, and is related to incompressibility.  A $D$-algebra on the other hand is related to volume preserving differomorphisms \cite{hoppe-1997-70}. Indeed it is a quantization of the classical Nambu-Poisson bracket \cite{nambu-1980}, which is known to be invariant under volume preserving diffeomorphisms. It would be interesting to make this connection more explicit and to understand its link to the incompressibility of TIs in higher dimensions.

 Moreover, the GMP algebra is related to a two-dimensional Aharonov-Bohm effect of pointlike objects moving in the background of the Berry curvature $F$. In higher dimensions, the $D$-algebra  involves the $D$-form $F\wedge \cdots \wedge F$.  The natural objects that can couple to a $D$-form are  $D-2$ dimensional membranes  \cite{baez-2011GReGr..43.2335B}. Interestingly, the classical limit of the $D$-commutator is the Nambu-Poisson bracket \cite{nambu-PhysRevD.7.2405}, which is a natural setup to describe the dynamics of classical membranes \cite{nambu-1980}. The appearance of extended objects in the field theory description of topological insulators in dimensions greater than three is also expected from the BF proposal of \cite{cho-2011AnPhy.326.1515C}. This suggests that the correct "effective" description of the higher-dimensional topological insulators is in terms of parallel transport not of electrons but of extended objects, such as strings in $3$ dimensions. We speculate the the Chern-simons term could appear when such algebras are constructed.

\emph{Note added:} We recently became aware of a related paper\cite{neupert-PhysRevB.86.035125}. While most of our results are similar, our conclusions  in odd space dimensions are exactly the opposite. We have shown that it is not possible to obtain the $\mathbb{Z}_2$ topological invariant through the algebra of the projected density operators.

\emph{Acknowledgments: } We wish to thank D.~Haldane, S.~Sondi, S.~Parameswaran, P.~Wiegmann, T.~Hughes, and S.~Ryu for fruitful discussions. BAB was supported by Princeton Startup Funds, NSF CAREER Grant No. DMR-095242, Grant ONR No. - N00014-11-1-0635, Darpa - N66001-11-1-4110, the Packard Foundation and a Keck grant. NR was supported by the Packard Foundation and a Keck grant. BE was supported by ONR Grant No. N00014-11-1-0635.

\bibliography{dalgebra.bib}

\begin{thebibliography}{34}
\expandafter\ifx\csname natexlab\endcsname\relax\def\natexlab#1{#1}\fi
\expandafter\ifx\csname bibnamefont\endcsname\relax
  \def\bibnamefont#1{#1}\fi
\expandafter\ifx\csname bibfnamefont\endcsname\relax
  \def\bibfnamefont#1{#1}\fi
\expandafter\ifx\csname citenamefont\endcsname\relax
  \def\citenamefont#1{#1}\fi
\expandafter\ifx\csname url\endcsname\relax
  \def\url#1{\texttt{#1}}\fi
\expandafter\ifx\csname urlprefix\endcsname\relax\def\urlprefix{URL }\fi
\providecommand{\bibinfo}[2]{#2}
\providecommand{\eprint}[2][]{\url{#2}}

\bibitem[{\citenamefont{Kane and Mele}(2005)}]{kane-PhysRevLett.95.226801}
\bibinfo{author}{\bibfnamefont{C.~L.} \bibnamefont{Kane}} \bibnamefont{and}
  \bibinfo{author}{\bibfnamefont{E.~J.} \bibnamefont{Mele}},
  \bibinfo{journal}{Phys. Rev. Lett.} \textbf{\bibinfo{volume}{95}},
  \bibinfo{pages}{226801} (\bibinfo{year}{2005}).

\bibitem[{\citenamefont{Bernevig et~al.}(2006)\citenamefont{Bernevig, Hughes,
  and Zhang}}]{Bernevig15122006}
\bibinfo{author}{\bibfnamefont{B.~A.} \bibnamefont{Bernevig}},
  \bibinfo{author}{\bibfnamefont{T.~L.} \bibnamefont{Hughes}},
  \bibnamefont{and} \bibinfo{author}{\bibfnamefont{S.-C.} \bibnamefont{Zhang}},
  \bibinfo{journal}{Science} \textbf{\bibinfo{volume}{314}},
  \bibinfo{pages}{1757} (\bibinfo{year}{2006}).

\bibitem[{\citenamefont{{Sheng} et~al.}(2011)\citenamefont{{Sheng}, {Gu},
  {Sun}, and {Sheng}}}]{sheng-natcommun.2.389}
\bibinfo{author}{\bibfnamefont{D.~N.} \bibnamefont{{Sheng}}},
  \bibinfo{author}{\bibfnamefont{Z.-C.} \bibnamefont{{Gu}}},
  \bibinfo{author}{\bibfnamefont{K.}~\bibnamefont{{Sun}}}, \bibnamefont{and}
  \bibinfo{author}{\bibfnamefont{L.}~\bibnamefont{{Sheng}}},
  \bibinfo{journal}{Nat Commun} \textbf{\bibinfo{volume}{2}},
  \bibinfo{pages}{389} (\bibinfo{year}{2011}).

\bibitem[{\citenamefont{Regnault and
  Bernevig}(2011)}]{regnault-PhysRevX.1.021014}
\bibinfo{author}{\bibfnamefont{N.}~\bibnamefont{Regnault}} \bibnamefont{and}
  \bibinfo{author}{\bibfnamefont{B.~A.} \bibnamefont{Bernevig}},
  \bibinfo{journal}{Phys. Rev. X} \textbf{\bibinfo{volume}{1}},
  \bibinfo{pages}{021014} (\bibinfo{year}{2011}).

\bibitem[{\citenamefont{Neupert
  et~al.}(2011{\natexlab{a}})\citenamefont{Neupert, Santos, Chamon, and
  Mudry}}]{neupert-PhysRevLett.106.236804}
\bibinfo{author}{\bibfnamefont{T.}~\bibnamefont{Neupert}},
  \bibinfo{author}{\bibfnamefont{L.}~\bibnamefont{Santos}},
  \bibinfo{author}{\bibfnamefont{C.}~\bibnamefont{Chamon}}, \bibnamefont{and}
  \bibinfo{author}{\bibfnamefont{C.}~\bibnamefont{Mudry}},
  \bibinfo{journal}{Phys. Rev. Lett.} \textbf{\bibinfo{volume}{106}},
  \bibinfo{pages}{236804} (\bibinfo{year}{2011}{\natexlab{a}}).

\bibitem[{\citenamefont{Wang et~al.}(2011)\citenamefont{Wang, Gu, Gong, and
  Sheng}}]{wang-PhysRevLett.107.146803}
\bibinfo{author}{\bibfnamefont{Y.-F.} \bibnamefont{Wang}},
  \bibinfo{author}{\bibfnamefont{Z.-C.} \bibnamefont{Gu}},
  \bibinfo{author}{\bibfnamefont{C.-D.} \bibnamefont{Gong}}, \bibnamefont{and}
  \bibinfo{author}{\bibfnamefont{D.~N.} \bibnamefont{Sheng}},
  \bibinfo{journal}{Phys. Rev. Lett.} \textbf{\bibinfo{volume}{107}},
  \bibinfo{pages}{146803} (\bibinfo{year}{2011}).

\bibitem[{\citenamefont{Wu et~al.}(2012)\citenamefont{Wu, Bernevig, and
  Regnault}}]{wu-PhysRevB.85.075116}
\bibinfo{author}{\bibfnamefont{Y.-L.} \bibnamefont{Wu}},
  \bibinfo{author}{\bibfnamefont{B.~A.} \bibnamefont{Bernevig}},
  \bibnamefont{and} \bibinfo{author}{\bibfnamefont{N.}~\bibnamefont{Regnault}},
  \bibinfo{journal}{Phys. Rev. B} \textbf{\bibinfo{volume}{85}},
  \bibinfo{pages}{075116} (\bibinfo{year}{2012}).

\bibitem[{\citenamefont{Qi}(2011)}]{qi-PhysRevLett.107.126803}
\bibinfo{author}{\bibfnamefont{X.-L.} \bibnamefont{Qi}},
  \bibinfo{journal}{Phys. Rev. Lett.} \textbf{\bibinfo{volume}{107}},
  \bibinfo{pages}{126803} (\bibinfo{year}{2011}).

\bibitem[{\citenamefont{Wang and Ran}(2011)}]{wang-PhysRevB.84.241103}
\bibinfo{author}{\bibfnamefont{F.}~\bibnamefont{Wang}} \bibnamefont{and}
  \bibinfo{author}{\bibfnamefont{Y.}~\bibnamefont{Ran}},
  \bibinfo{journal}{Phys. Rev. B} \textbf{\bibinfo{volume}{84}},
  \bibinfo{pages}{241103} (\bibinfo{year}{2011}).

\bibitem[{\citenamefont{McGreevy et~al.}(2012)\citenamefont{McGreevy, Swingle,
  and Tran}}]{McGreevy-PhysRevB.85.125105}
\bibinfo{author}{\bibfnamefont{J.}~\bibnamefont{McGreevy}},
  \bibinfo{author}{\bibfnamefont{B.}~\bibnamefont{Swingle}}, \bibnamefont{and}
  \bibinfo{author}{\bibfnamefont{K.-A.} \bibnamefont{Tran}},
  \bibinfo{journal}{Phys. Rev. B} \textbf{\bibinfo{volume}{85}},
  \bibinfo{pages}{125105} (\bibinfo{year}{2012}).

\bibitem[{\citenamefont{Jiang et~al.}(2011)\citenamefont{Jiang, Lu, Zhai, Low,
  and Hu}}]{jiang-PhysRevB.84.205324}
\bibinfo{author}{\bibfnamefont{Y.}~\bibnamefont{Jiang}},
  \bibinfo{author}{\bibfnamefont{F.}~\bibnamefont{Lu}},
  \bibinfo{author}{\bibfnamefont{F.}~\bibnamefont{Zhai}},
  \bibinfo{author}{\bibfnamefont{T.}~\bibnamefont{Low}}, \bibnamefont{and}
  \bibinfo{author}{\bibfnamefont{J.}~\bibnamefont{Hu}}, \bibinfo{journal}{Phys.
  Rev. B} \textbf{\bibinfo{volume}{84}}, \bibinfo{pages}{205324}
  (\bibinfo{year}{2011}).

\bibitem[{\citenamefont{Neupert
  et~al.}(2011{\natexlab{b}})\citenamefont{Neupert, Santos, Ryu, Chamon, and
  Mudry}}]{neupert-PhysRevB.84.165107}
\bibinfo{author}{\bibfnamefont{T.}~\bibnamefont{Neupert}},
  \bibinfo{author}{\bibfnamefont{L.}~\bibnamefont{Santos}},
  \bibinfo{author}{\bibfnamefont{S.}~\bibnamefont{Ryu}},
  \bibinfo{author}{\bibfnamefont{C.}~\bibnamefont{Chamon}}, \bibnamefont{and}
  \bibinfo{author}{\bibfnamefont{C.}~\bibnamefont{Mudry}},
  \bibinfo{journal}{Phys. Rev. B} \textbf{\bibinfo{volume}{84}},
  \bibinfo{pages}{165107} (\bibinfo{year}{2011}{\natexlab{b}}).

\bibitem[{\citenamefont{Santos et~al.}(2011)\citenamefont{Santos, Neupert, Ryu,
  Chamon, and Mudry}}]{neupert-PhysRevB.84.165138}
\bibinfo{author}{\bibfnamefont{L.}~\bibnamefont{Santos}},
  \bibinfo{author}{\bibfnamefont{T.}~\bibnamefont{Neupert}},
  \bibinfo{author}{\bibfnamefont{S.}~\bibnamefont{Ryu}},
  \bibinfo{author}{\bibfnamefont{C.}~\bibnamefont{Chamon}}, \bibnamefont{and}
  \bibinfo{author}{\bibfnamefont{C.}~\bibnamefont{Mudry}},
  \bibinfo{journal}{Phys. Rev. B} \textbf{\bibinfo{volume}{84}},
  \bibinfo{pages}{165138} (\bibinfo{year}{2011}).

\bibitem[{\citenamefont{Hugues et~al.}()\citenamefont{Hugues, Bernevig, and
  Regnault}}]{Taylor-inprep2012}
\bibinfo{author}{\bibfnamefont{T.}~\bibnamefont{Hugues}},
  \bibinfo{author}{\bibfnamefont{B.~A.} \bibnamefont{Bernevig}},
  \bibnamefont{and} \bibinfo{author}{\bibfnamefont{N.}~\bibnamefont{Regnault}},
  \bibinfo{note}{in preparation}.

\bibitem[{\citenamefont{Bernevig and
  Regnault}(2012)}]{Bernevig-2012PhysRevB.85.075128}
\bibinfo{author}{\bibfnamefont{B.~A.} \bibnamefont{Bernevig}} \bibnamefont{and}
  \bibinfo{author}{\bibfnamefont{N.}~\bibnamefont{Regnault}},
  \bibinfo{journal}{Phys. Rev. B} \textbf{\bibinfo{volume}{85}},
  \bibinfo{pages}{075128} (\bibinfo{year}{2012}).

\bibitem[{\citenamefont{Parameswaran et~al.}(2012)\citenamefont{Parameswaran,
  Roy, and Sondhi}}]{Parameswaran-PhysRevB.85.241308}
\bibinfo{author}{\bibfnamefont{S.~A.} \bibnamefont{Parameswaran}},
  \bibinfo{author}{\bibfnamefont{R.}~\bibnamefont{Roy}}, \bibnamefont{and}
  \bibinfo{author}{\bibfnamefont{S.~L.} \bibnamefont{Sondhi}},
  \bibinfo{journal}{Phys. Rev. B} \textbf{\bibinfo{volume}{85}},
  \bibinfo{pages}{241308} (\bibinfo{year}{2012}).

\bibitem[{\citenamefont{Girvin et~al.}(1986)\citenamefont{Girvin, MacDonald,
  and Platzman}}]{girvin-PhysRevB.33.2481}
\bibinfo{author}{\bibfnamefont{S.~M.} \bibnamefont{Girvin}},
  \bibinfo{author}{\bibfnamefont{A.~H.} \bibnamefont{MacDonald}},
  \bibnamefont{and} \bibinfo{author}{\bibfnamefont{P.~M.}
  \bibnamefont{Platzman}}, \bibinfo{journal}{Phys. Rev. B}
  \textbf{\bibinfo{volume}{33}}, \bibinfo{pages}{2481} (\bibinfo{year}{1986}).

\bibitem[{\citenamefont{Wiegmann}(2012)}]{Wiegmann-PhysRevLett.108.206810}
\bibinfo{author}{\bibfnamefont{P.}~\bibnamefont{Wiegmann}},
  \bibinfo{journal}{Phys. Rev. Lett.} \textbf{\bibinfo{volume}{108}},
  \bibinfo{pages}{206810} (\bibinfo{year}{2012}).

\bibitem[{\citenamefont{{Kitaev}}(2009)}]{Kitaev-2009AIPC.1134.22K}
\bibinfo{author}{\bibfnamefont{A.}~\bibnamefont{{Kitaev}}}, in
  \emph{\bibinfo{booktitle}{American Institute of Physics Conference Series}},
  edited by \bibinfo{editor}{\bibfnamefont{V.}~\bibnamefont{{Lebedev}}}
  \bibnamefont{and}
  \bibinfo{editor}{\bibfnamefont{M.}~\bibnamefont{{Feigel'Man}}}
  (\bibinfo{year}{2009}), vol. \bibinfo{volume}{1134} of
  \emph{\bibinfo{series}{American Institute of Physics Conference Series}}, pp.
  \bibinfo{pages}{22--30}, \eprint{0901.2686}.

\bibitem[{\citenamefont{{Ryu} et~al.}(2010)\citenamefont{{Ryu}, {Schnyder},
  {Furusaki}, and {Ludwig}}}]{Ryu-2010NJPh.12f5010R}
\bibinfo{author}{\bibfnamefont{S.}~\bibnamefont{{Ryu}}},
  \bibinfo{author}{\bibfnamefont{A.~P.} \bibnamefont{{Schnyder}}},
  \bibinfo{author}{\bibfnamefont{A.}~\bibnamefont{{Furusaki}}},
  \bibnamefont{and} \bibinfo{author}{\bibfnamefont{A.~W.~W.}
  \bibnamefont{{Ludwig}}}, \bibinfo{journal}{New Journal of Physics}
  \textbf{\bibinfo{volume}{12}}, \bibinfo{pages}{065010}
  (\bibinfo{year}{2010}).

\bibitem[{\citenamefont{{Goerbig, M.O.}}(2012)}]{goerbig-2012epjb}
\bibinfo{author}{\bibnamefont{{Goerbig, M.O.}}}, \bibinfo{journal}{Eur. Phys.
  J. B} \textbf{\bibinfo{volume}{85}}, \bibinfo{pages}{15}
  (\bibinfo{year}{2012}).

\bibitem[{\citenamefont{{Podolsky} and
  {Avron}}()}]{podolskiandavronprivatecommunication}
\bibinfo{author}{\bibfnamefont{D.}~\bibnamefont{{Podolsky}}} \bibnamefont{and}
  \bibinfo{author}{\bibfnamefont{J.}~\bibnamefont{{Avron}}},
  \bibinfo{note}{private communication}.

\bibitem[{\citenamefont{{Zhang} and {Hu}}(2001)}]{Zhang-2001Sci.294.823Z}
\bibinfo{author}{\bibfnamefont{S.-C.} \bibnamefont{{Zhang}}} \bibnamefont{and}
  \bibinfo{author}{\bibfnamefont{J.}~\bibnamefont{{Hu}}},
  \bibinfo{journal}{Science} \textbf{\bibinfo{volume}{294}},
  \bibinfo{pages}{823} (\bibinfo{year}{2001}).

\bibitem[{\citenamefont{Bernevig et~al.}(2003)\citenamefont{Bernevig, Hu,
  Toumbas, and Zhang}}]{PhysRevLett.91.236803}
\bibinfo{author}{\bibfnamefont{B.~A.} \bibnamefont{Bernevig}},
  \bibinfo{author}{\bibfnamefont{J.}~\bibnamefont{Hu}},
  \bibinfo{author}{\bibfnamefont{N.}~\bibnamefont{Toumbas}}, \bibnamefont{and}
  \bibinfo{author}{\bibfnamefont{S.-C.} \bibnamefont{Zhang}},
  \bibinfo{journal}{Phys. Rev. Lett.} \textbf{\bibinfo{volume}{91}},
  \bibinfo{pages}{236803} (\bibinfo{year}{2003}).

\bibitem[{\citenamefont{Dunne and Trugenberger}(1990)}]{Dunne1990281}
\bibinfo{author}{\bibfnamefont{G.~V.} \bibnamefont{Dunne}} \bibnamefont{and}
  \bibinfo{author}{\bibfnamefont{C.~A.} \bibnamefont{Trugenberger}},
  \bibinfo{journal}{Annals of Physics} \textbf{\bibinfo{volume}{204}},
  \bibinfo{pages}{281 } (\bibinfo{year}{1990}), ISSN \bibinfo{issn}{0003-4916}.

\bibitem[{\citenamefont{Curtright and
  Zachos}(2003)}]{Curtright-PhysRevD.68.085001}
\bibinfo{author}{\bibfnamefont{T.}~\bibnamefont{Curtright}} \bibnamefont{and}
  \bibinfo{author}{\bibfnamefont{C.}~\bibnamefont{Zachos}},
  \bibinfo{journal}{Phys. Rev. D} \textbf{\bibinfo{volume}{68}},
  \bibinfo{pages}{085001} (\bibinfo{year}{2003}).

\bibitem[{\citenamefont{Baez et~al.}(2010)\citenamefont{Baez, Hoffnung, and
  Rogers}}]{baez-commmathphys.293.2010}
\bibinfo{author}{\bibfnamefont{J.}~\bibnamefont{Baez}},
  \bibinfo{author}{\bibfnamefont{A.}~\bibnamefont{Hoffnung}}, \bibnamefont{and}
  \bibinfo{author}{\bibfnamefont{C.}~\bibnamefont{Rogers}},
  \bibinfo{journal}{Communications in Mathematical Physics}
  \textbf{\bibinfo{volume}{293}}, \bibinfo{pages}{701} (\bibinfo{year}{2010}),
  ISSN \bibinfo{issn}{0010-3616}, \bibinfo{note}{10.1007/s00220-009-0951-9}.

\bibitem[{\citenamefont{{Baez} and {Huerta}}(2011)}]{baez-2011GReGr..43.2335B}
\bibinfo{author}{\bibfnamefont{J.~C.} \bibnamefont{{Baez}}} \bibnamefont{and}
  \bibinfo{author}{\bibfnamefont{J.}~\bibnamefont{{Huerta}}},
  \bibinfo{journal}{General Relativity and Gravitation}
  \textbf{\bibinfo{volume}{43}}, \bibinfo{pages}{2335} (\bibinfo{year}{2011}),
  \eprint{1003.4485}.

\bibitem[{\citenamefont{{Cho} and {Moore}}(2011)}]{cho-2011AnPhy.326.1515C}
\bibinfo{author}{\bibfnamefont{G.~Y.} \bibnamefont{{Cho}}} \bibnamefont{and}
  \bibinfo{author}{\bibfnamefont{J.~E.} \bibnamefont{{Moore}}},
  \bibinfo{journal}{Annals of Physics} \textbf{\bibinfo{volume}{326}},
  \bibinfo{pages}{1515} (\bibinfo{year}{2011}), \eprint{1011.3485}.

\bibitem[{\citenamefont{Horowitz}(1989)}]{Horowitz-CommMathPhys.1989}
\bibinfo{author}{\bibfnamefont{G.~T.} \bibnamefont{Horowitz}},
  \bibinfo{journal}{Communications in Mathematical Physics}
  \textbf{\bibinfo{volume}{125}}, \bibinfo{pages}{417} (\bibinfo{year}{1989}),
  ISSN \bibinfo{issn}{0010-3616}, \bibinfo{note}{10.1007/BF01218410}.

\bibitem[{\citenamefont{Hoppe}(1997)}]{hoppe-1997-70}
\bibinfo{author}{\bibfnamefont{J.}~\bibnamefont{Hoppe}},
  \bibinfo{journal}{HELV.PHYS.ACTA} \textbf{\bibinfo{volume}{70}},
  \bibinfo{pages}{302} (\bibinfo{year}{1997}).

\bibitem[{\citenamefont{Y. and Nambu}(1980)}]{nambu-1980}
\bibinfo{author}{\bibnamefont{Y.}} \bibnamefont{and}
  \bibinfo{author}{\bibnamefont{Nambu}}, \bibinfo{journal}{Physics Letters B}
  \textbf{\bibinfo{volume}{92}}, \bibinfo{pages}{327 } (\bibinfo{year}{1980}),
  ISSN \bibinfo{issn}{0370-2693}.

\bibitem[{\citenamefont{Nambu}(1973)}]{nambu-PhysRevD.7.2405}
\bibinfo{author}{\bibfnamefont{Y.}~\bibnamefont{Nambu}},
  \bibinfo{journal}{Phys. Rev. D} \textbf{\bibinfo{volume}{7}},
  \bibinfo{pages}{2405} (\bibinfo{year}{1973}).

\bibitem[{\citenamefont{Neupert et~al.}(2012)\citenamefont{Neupert, Santos,
  Ryu, Chamon, and Mudry}}]{neupert-PhysRevB.86.035125}
\bibinfo{author}{\bibfnamefont{T.}~\bibnamefont{Neupert}},
  \bibinfo{author}{\bibfnamefont{L.}~\bibnamefont{Santos}},
  \bibinfo{author}{\bibfnamefont{S.}~\bibnamefont{Ryu}},
  \bibinfo{author}{\bibfnamefont{C.}~\bibnamefont{Chamon}}, \bibnamefont{and}
  \bibinfo{author}{\bibfnamefont{C.}~\bibnamefont{Mudry}},
  \bibinfo{journal}{Phys. Rev. B} \textbf{\bibinfo{volume}{86}},
  \bibinfo{pages}{035125} (\bibinfo{year}{2012}).

\end{thebibliography}

\end{document}